\documentclass[]{spie}  

\usepackage{amsmath,amsfonts,amssymb}
\usepackage{rotating,caption,graphicx,subfig}
\usepackage[colorlinks=true, allcolors=blue]{hyperref}

\title{HIRAX: A Probe of Dark Energy and Radio Transients}

\author[a]{L.B. Newburgh}
\author[b,c]{K. Bandura}
\author[d,e]{M. A. Bucher}
\author[f]{T.-C. Chang}
\author[g,h]{H.C. Chiang}
\author[i]{J.F. Cliche}
\author[j,k,l]{R. Dav\'e}
\author[i]{M. Dobbs}
\author[m,n]{C. Clarkson}
\author[d]{K. M. Ganga}
\author[e]{T. Gogo}
\author[o]{A. Gumba}
\author[p]{N. Gupta}
\author[h]{M. Hilton}
\author[b,c]{B. Johnstone}
\author[q,r,s]{A. Karastergiou}
\author[t]{M. Kunz}
\author[a,u]{D. Lokhorst}
\author[q]{R. Maartens}
\author[o]{S. Macpherson}
\author[e]{M. Mdlalose}
\author[h]{K. Moodley}
\author[e]{L. Ngwenya}
\author[i]{J.M. Parra}
\author[v]{J. Peterson}
\author[a]{O. Recnik}
\author[h]{B. Saliwanchik}
\author[q]{M. G. Santos}
\author[e,g]{J.L. Sievers}
\author[s]{O. Smirnov}
\author[w]{P. Stronkhorst}
\author[m]{R. Taylor}
\author[a,u]{K. Vanderlinde}
\author[o]{G. Van~Vuuren}
\author[m,x,y]{A. Weltman}
\author[q]{A. Witzemann}

\affil[a]{Dunlap Institute, University of Toronto, 50 St. George St., Toronto, Canada }
\affil[b]{Lane Department of Computer Science and Electrical Engineering, West Virginia University, PO Box 6201, Morgantown, WV 26506, USA}
\affil[c]{Department of Physics and Astronomy, West Virginia University, PO Box 6201, Morgantown, WV 26506, USA}
\affil[d]{APC, Univ Paris Diderot, CNRS/IN2P3, CEA/lrfu, Obs de Paris, Sorbonne Paris Cit\'e, France}
\affil[e]{Astrophysics \& Cosmology Research Unit, School of Chemistry and Physics, University of KwaZulu-Natal, Durban, South Africa National Institute for Theoretical Physics, KwaZulu-Natal, South Africa}
\affil[f]{Academia Sinica Institute of Astronomy and Astrophysics, 11F of ASMAB, AS/NTU, 1 Roosevelt Rd Sec. 4, Taipei, 10617, Taiwan}
\affil[g]{National Institute for Theoretical Physics (NITheP), KwaZulu-Natal, South Africa}
\affil[h]{Astrophysics \& Cosmology Research Unit, School of Mathematics, Statistics \& Computer Science, University of KwaZulu-Natal, Westville Campus, Durban 4000, South Africa}
\affil[i]{Department of Physics, McGill University, Montreal, Quebec H3A 2T8, Canada}
\affil[j]{University of the Western Cape, Bellville, Cape Town 7535, South Africa}
\affil[k]{South African Astronomical Observatories, Observatory, Cape Town 7925, South Africa}
\affil[l]{African Institute for Mathematical Sciences, Muizenberg, Cape Town 7945, South Africa}
\affil[m]{The Cosmology \& Gravity Group, Department of Mathematics and Applied Mathematics, University of Cape Town, Private Bag, Rondebosch, 7700, South Africa.}
\affil[n]{School of Physics \& Astronomy, Queen Mary University of London, Mile End Road, London E1 4NS, UK.}
\affil[o]{Durban University of Technology, Engineering and the Built Environment, P.O. Box 1334, Durban, 4000 South Africa}
\affil[p]{IUCAA, Post Bag 4, Ganeshkhind, Pune 411007, India}
\affil[q]{Department of Physics, University of the Western Cape, Cape Town 7535, South Africa and SKA SA, The Park, Park Road, Pinelands 7405, South Africa}
\affil[r]{Astrophysics, University of Oxford, Denys Wilkinson Building, Keble Road, Oxford OX1 3RH, UK}
\affil[s]{Department of Physics and Electronics, Rhodes University, PO Box 94, Grahamstown 6140, South Africa}
\affil[t]{Departement de Physique Theorique and Center for Astroparticle Physics, Universite de Geneve, 24 quai Ernest Ansermet, CH–1211 Geneve 4, Switzerland}
\affil[u]{Department of Astronomy and Astrophysics, University of Toronto, 50 St. George St., Toronto, Canada}
\affil[v]{McWilliams Center for Cosmology, Department of Physics, Carnegie Mellon University, 5000 Forbes Ave, Pittsburgh PA 15213, USA}
\affil[w]{HartRAO, P. O. Box 443, Krugersdorp 1740, South Africa}
\affil[x]{Department of Astrophysical Sciences, Princeton University, Peyton Hall, Princeton NJ 08544-0010, USA}
\affil[y]{School of Natural Sciences, Institute for Advanced Study, Olden Lane, Princeton, NJ 08540, USA}

\authorinfo{Further author information: (Send correspondence to L.B.N.)\\L.B.N.: E-mail: newburgh@dunlap.utoronto.ca}

\begin{document} 
\maketitle

\begin{abstract}
The Hydrogen Intensity and Real-time Analysis eXperiment (HIRAX) is a new 400--800\,MHz radio interferometer under development for deployment in South Africa. HIRAX will comprise 1024 six\,meter parabolic dishes on a compact grid and will map most of the southern sky over the course of four years. HIRAX has two primary science goals: to constrain Dark Energy and measure structure at high redshift, and to study radio transients and pulsars. HIRAX will observe unresolved sources of neutral hydrogen via their redshifted 21-cm emission line (`hydrogen intensity mapping'). The resulting maps of large-scale structure at redshifts 0.8--2.5 will be used to measure Baryon Acoustic Oscillations (BAO). BAO are a preferential length scale in the matter distribution that can be used to characterize the expansion history of the Universe and thus understand the properties of Dark Energy. HIRAX will improve upon current BAO measurements from galaxy surveys by observing a larger cosmological volume (larger in both survey area and redshift range) and by measuring BAO at higher redshift when the expansion of the universe transitioned to Dark Energy domination. HIRAX will complement CHIME, a hydrogen intensity mapping experiment in the Northern Hemisphere, by completing the sky coverage in the same redshift range. HIRAX's location in the Southern Hemisphere also allows a variety of cross-correlation measurements with large-scale structure surveys at many wavelengths. Daily maps of a few thousand square degrees of the Southern Hemisphere, encompassing much of the Milky Way galaxy, will also open new opportunities for discovering and monitoring radio transients. The HIRAX correlator will have the ability to rapidly and efficiently detect transient events. This new data will shed light on the poorly understood nature of fast radio bursts (FRBs), enable pulsar monitoring to enhance long-wavelength gravitational wave searches, and provide a rich data set for new radio transient phenomena searches. This paper discusses the HIRAX instrument, science goals, and current status.
\end{abstract}

\keywords{Cosmology, Dark Energy, large-scale Structure, Intensity Mapping, 21cm}

\section{INTRODUCTION}
\label{sec:intro}  

Measurements from Type Ia supernovae (SN1a)~\cite{SN1a}, Baryon Acoustic Oscillations (BAO)~\cite{2016MNRAS.457.1770C}, and the Cosmic Microwave Background (CMB)~\cite{2015arXiv150201589P} have shown that the energy density of the Universe is now dominated by Dark Energy, an unknown component causing the expansion rate of the Universe to increase. To better understand the nature of Dark Energy, we require measurements to a redshift of $z\sim2$, when Dark Energy began to influence the expansion rate. BAO provide a unique observational tool that can be extended to higher redshifts. BAO are a characteristic scale of 150\,Mpc in the matter power spectrum that is initially imprinted by primordial acoustic oscillations in the photon-–baryon fluid at decoupling.  Because large-scale structure preferentially forms at that co-moving 150\,Mpc scale, measurements of the large-scale structure, and hence the BAO length scale, at various redshifts are sensitive tracers of the expansion rate of the Universe, allowing us to probe Dark Energy and its evolution.  Recent galaxy surveys have already demonstrated percent-level constraints on Dark Energy at redshift $z\sim0.6$~\cite{2014MNRAS.441...24A} and have made a $5\sigma$ detection of high-redshift BAO using $\mathrm{Ly}\alpha$ forest measurements with quasars~\cite{2015A&A...574A..59D}.\newline

A promising technique for measuring BAO at higher redshifts is 21-cm intensity mapping, whereby galaxies are observed in aggregate through low-resolution measurements of red-shifted 21-cm emission of neutral hydrogen. This method has two benefits: the 21\,cm emission line provides a natural redshift marker, and we can focus sensitivity on the large scales of interest to trace redshift evolution without resolving individual galaxies.~\cite{2015PhRvD..91h3514S}~\cite{2015aska.confE..19S}~\cite{2015ApJ...803...21B} Currently, 21\,cm intensity mapping has been measured only in the cross-correlation between radio surveys and galaxy surveys at redshift $z\sim0.8$: a $\sim$4$\sigma$ detection with the DEEP2 galaxy survey~\cite{2010arXiv1007.3709C}, and at higher significance with WiggleZ~\cite{2013ApJ...763L..20M}. Only upper bounds have been placed on the BAO contribution to the 21\,cm auto-correlation power spectrum.~\cite{2013MNRAS.434L..46S} \newline

 \begin{figure} [t]
   \begin{center}
   \begin{tabular}{c} 
  \includegraphics[height=7cm]{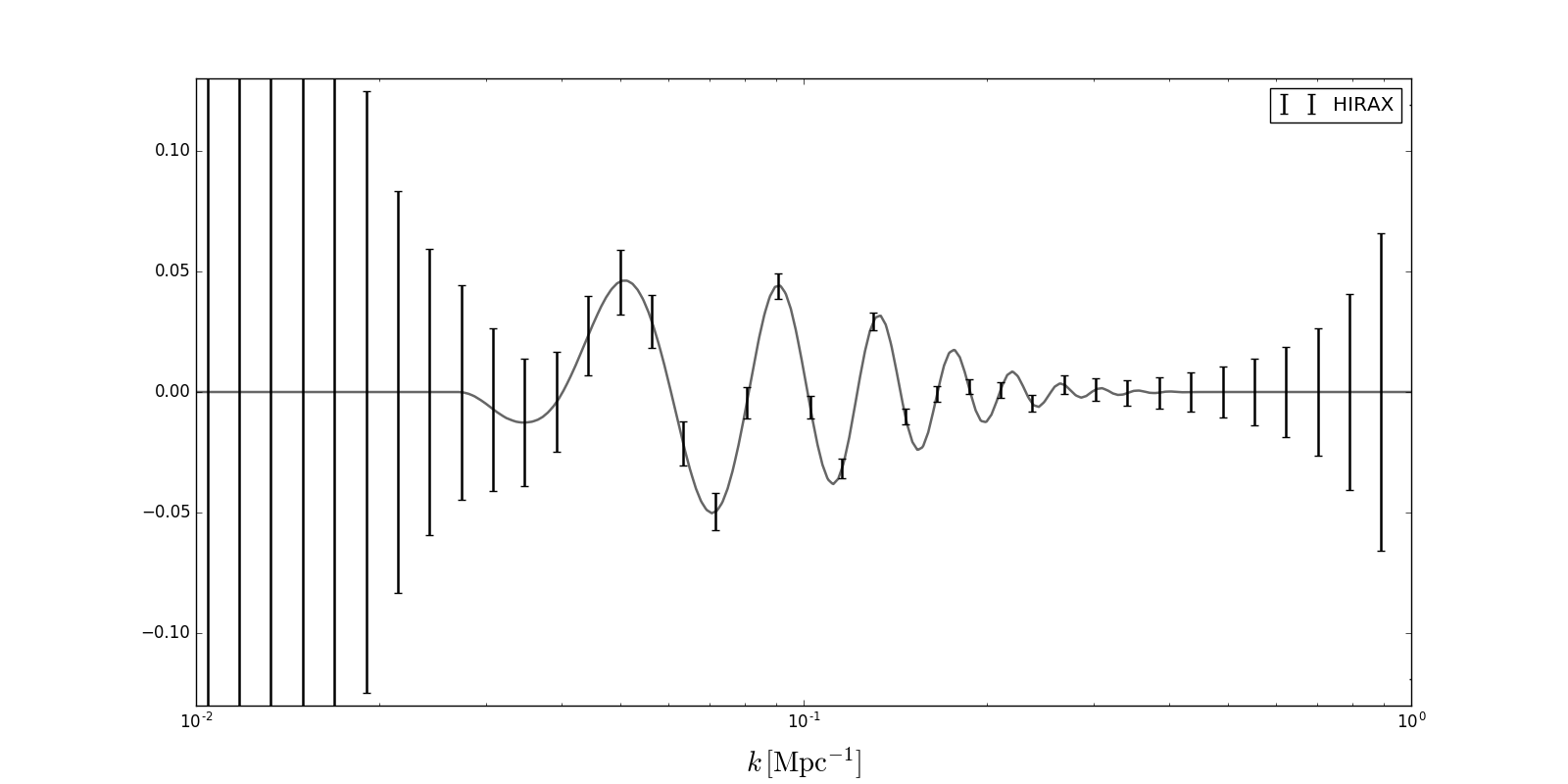}
   \end{tabular}
   \end{center}
   \caption[dish] 
   { \label{fig:pspec} 
Expected baryon acoustic oscillations measurements with HIRAX, assuming a total sky area of 2,000 square degrees, 50\,K system noise, 1024 6\,m dishes, and 10,000 hours of integrated time, with Planck priors. We have divided by a fiducial model to make the power spectrum features more apparent. No systematics have been assumed. }
   \end{figure}

\begin{figure} [t]
   \begin{center}
   \begin{tabular}{c} 
  \includegraphics[height=9cm]{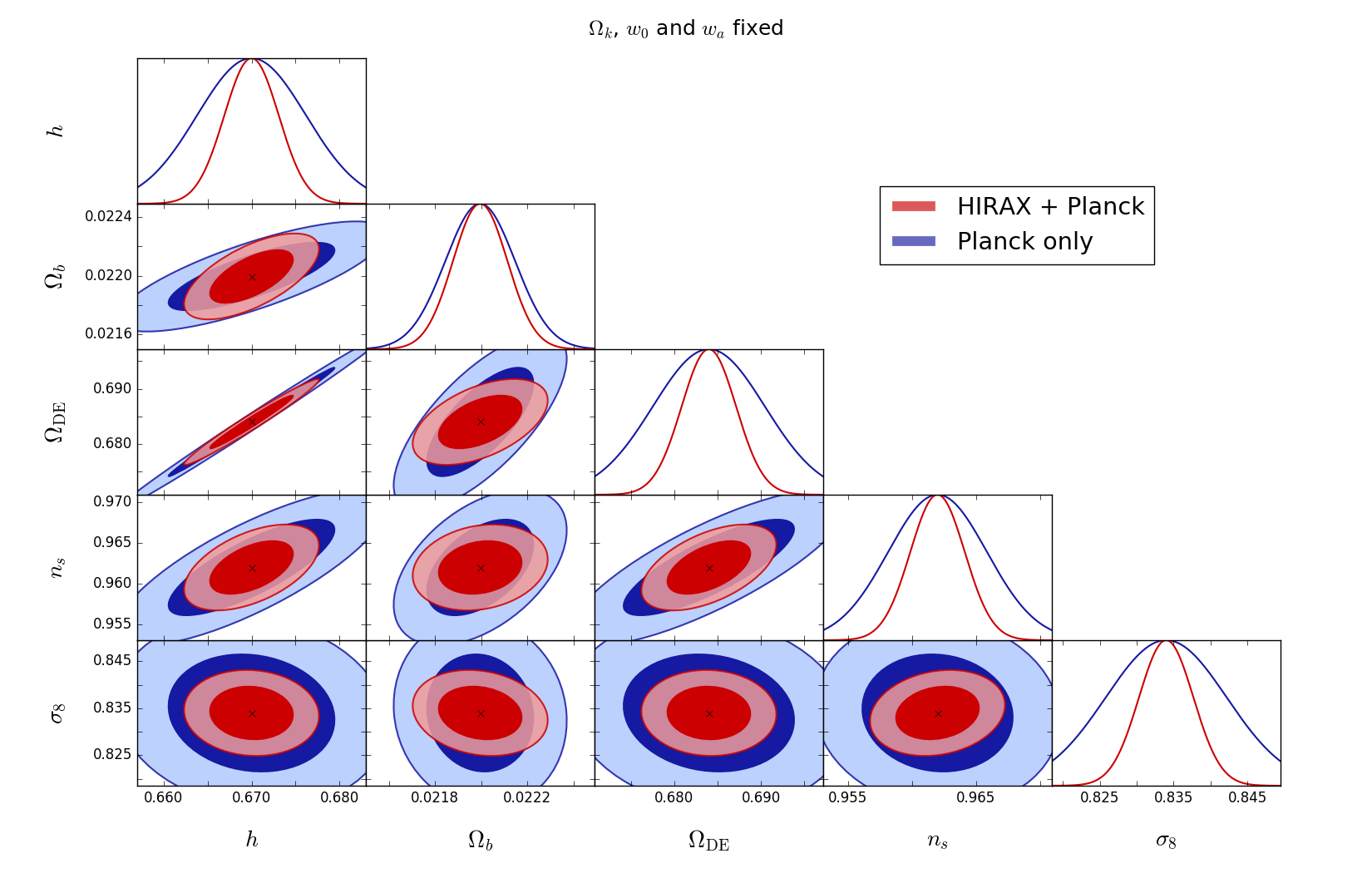}
   \end{tabular}
   \end{center}
   \caption[dish] 
   { \label{fig:params} 
Forecast of cosmological parameter constraints for HIRAX assuming a total sky area of 2,000 square degrees, 50\,K system noise, 1024 6\,m dishes, and 10,000 hours of integrated time, with Planck priors.}
   \end{figure}

The Hydrogen Intensity and Real-time Analysis eXperiment (HIRAX)\footnote{http:\//\//www.acru.ukzn.ac.za\//$\sim$hirax} is a radio interferometer being developed for observations in South Africa operating in the frequency range 400-800\,MHz. HIRAX will map large-scale structure in a redshift range of $0.8 < z < 2.5$ with 1024 six\,meter dishes. HIRAX will observe $\sim$15,000 square degrees in the Southern Hemisphere, complementing CHIME \cite{2014SPIE.9145E..22B} in the Northern Hemisphere. The resulting maps of large-scale structure will measure the first four peaks in the BAO matter power spectrum, reaching sample variance limits, as shown in Figure~\ref{fig:pspec}, and improve constraints on cosmological parameters relating to structure at late times (the amplitude of the matter power spectrum, $\sigma_{8}$; total baryon density, $\Omega_{b}$; total Dark Energy density $\Omega_{DE}$; and the power spectrum spectral index, $n_{s}$) to less than 1\% (Figure~\ref{fig:params}). Observations from experiments that span a wide range of redshifts, like HIRAX, allow us to measure both geometry and growth, provide a probe of dynamical Dark Energy, constrain modified gravity models, measure the isotropy of the Universe, and improve limits on deviations from Gaussianity of the initial density perturbations (See for example discussion in Ref.~\cite{2015aska.confE..19S}). Constraints on cosmological parameters require end-to-end simulations linking the underlying dark matter distribution with HI emission. In addition, these simulations will use combinations of analytical models, hydrodynamical N-body code (See Ref.~\cite{2016arXiv160401418D}) and treatment of instrumental effects and of polarized foregrounds (See Ref.~\cite{2014MNRAS.444.3183A}).  \newline

We will take advantage of the Southern location to overlap with many surveys at other wavelengths for cross-correlation science, including both ongoing surveys (ACTPol~\cite{2016arXiv160506569T}, SPTpol~\cite{2014SPIE.9153E..1PB}, SDSS BOSS~\cite{2013AJ....145...10D},DES~\cite{2016MNRAS.460.1270D}, HST~\cite{2014SPIE.9147E..5QW}) and future surveys (Advanced ACT\cite{2016JLTP..tmp..144H}, SPT-3G\cite{2014SPIE.9153E..1PB},DESI~\cite{2013arXiv1308.0847L}, LSST~\cite{2012arXiv1211.0310L}, Euclid~\cite{2012SPIE.8442E..0TL}, WFIRST~\cite{2013arXiv1305.5422S}). Cross-correlations with optical galaxy redshift surveys and cosmic shear surveys will provide measurements of the redshift-dependent neutral hydrogen fraction ($\Omega_{\mathrm{HI}}$) and bias ($\mathrm{b}_{\mathrm{HI}}$), both of which are poorly constrained at these redshifts, and probe the relationship between stars and gas in their dark matter halos. We can also use any combination of optical and neutral hydrogen galaxy distributions to remove cosmic variance, dramatically improving constraints in the future (See for example Refs. ~\cite{2009JCAP...10..007M}~\cite{2013MNRAS.432..318A}). These cross-correlations, including millimeter experiments like ACT and SPT, will also help identify and remove systematics between the different surveys and potentially allow intensity mapping surveys to better characterize the foregrounds and optimize their removal. Direct correlation with CMB surveys is challenging because of the loss of long-wavelength line-of-sight modes in the 21cm density field after foreground removal (see Section~\ref{sec:chall}), however, higher order correlations of the 21cm density field~\cite{moodley} or tidal field reconstruction\cite{2016PhRvD..93j3504Z} may provide an observable signal, allowing us to constrain the high-redshift matter power spectrum as well as $\Omega_{\mathrm{HI}}$ and $b_{\mathrm{HI}}$ independently of the optical galaxy bias.

Observations of the southern sky provide access to the Galactic plane, enabling a wide variety of transient measurements with HIRAX. A new window on the transient sky will be possible in the near future from upcoming wide-field imagers like LSST and gravitational wave detections of coalescing compact objects from LIGO~\cite{PhysRevLett.116.241103} and explosions. HIRAX will add radio transient monitoring to this suite of observations, which will open up the discovery space for transient phenomena and contribute to multi-messenger science, for example following up nearby explosive events found by LIGO~\cite{2013MNRAS.430.2121P}. In addition, Fast Radio Bursts (FRBs) are a source of enormous interest to the radio transient community because of their high dispersion measures and isotropic spatial distributions \cite{2016MNRAS.460.1054C}, indicating possible cosmological distances. Their origin is unknown and the subject of ongoing study (See for example Ref.~\cite{2015Natur.528..523M}). Projecting from current discovery rates of FRBs, HIRAX will find dozens per day (with significant uncertainty in the estimated number due to limited FRB statistics in the HIRAX band) and be able to measure properties associated with their spectra, pulse arrival times, and spatial distribution. In addition to FRBs, HIRAX can be used as a pulsar discovery engine and an efficient pulsar monitoring telescope. HIRAX pulsar searches have the potential to discover a few thousand new pulsars, while pulsar monitoring would provide important inputs to timing arrays used for the detection of long-wavelength gravitational waves, inaccessible to other probes.\cite{2010CQGra..27h4013H}. Finally, up-channelizing the data to a spectral resolution of $\sim$3 kHz (1.5\,km/s in the center of the band) will yield a spectral rms of $\sim$4.4\,mJy/beam. This will provide a competitive 21\,cm absorption line survey in a redshift range $1.36<\mathrm{z}<2.5$ which will not be covered by any of the upcoming radio surveys (MeerKAT\footnote{http:\//\//www.ska.ac.za\//meerkat}, ASKAP\cite{2012SPIE.8444E..2AS} and APERTIF\footnote{http:\//\//www.astron.nl\//general\//apertif\//apertif}). \newline 

In this paper we describe the HIRAX instrument (Section~\ref{sec:instru}) including its reflector design, analog chain, and digitization. Foreground removal and calibration challenges  are described in Section~\ref{sec:chall}.

\section{The HIRAX Instrument}
\label{sec:instru}

The main driver for the HIRAX interferometer design is the measurement of large-scale structure at high redshift. The redshift-dependent 150\,Mpc BAO feature ranges in angular scale from 1.35$^{\circ}$ at $z\sim$2.5 (400\,MHz) to 3$^{\circ}$ at $z\sim$0.8 (800\,MHz), requiring a minimum baseline of $\sim$40\,m to resolve just the first peak, and a frequency resolution of 12\,MHz to resolve the BAO feature along the line of sight (in redshift). The signal level is also small, $\mathcal{O}$(0.1\,mK), requiring a low system noise and a large collecting area. Finally, we must reject galactic foregrounds that are nearly six orders of magnitude brighter than our cosmological signal of interest. This places stringent requirements on instrument calibration (described in Section~\ref{sec:chall}). In addition, to boost our sensitivity to the scales of interest and to keep the data storage and analysis tractable, we will design HIRAX to have a high degree of geometric redundancy. To take advantage of that redundancy, we must have \,cm-level repeatability in the baseline spacing and similarly tight requirements on beam conformity between dish elements. The HIRAX design (described below) has been optimized to meet these requirements. \newline 

The HIRAX instrument will be comprised of 1024 six\,meter dishes deployed in a 32$\times$32 grid ($\sim$250\,m$\times$250\,m) with the square sides aligned on the celestial cardinal directions. The signal chain is shown in Figure~\ref{fig:instrument} for each of the 1024 dishes: one radio dish, one dual-polarization antenna feed, two amplifiers (one per polarization), and Radio-Frequency-over-Fiber (RFoF) to carry the signal to the correlator building. The correlator is composed of a set of field-programmable gate array (FPGA) based digitizer/channelizer boards and a HPC graphics-processing unit (GPU) computer cluster for the spatial correlation, providing 1024 frequency bins in the 400--800\,MHz bandwidth, corresponding to a channelization of 390\,kHz. HIRAX is a transit telescope: all dishes will be pointed at the meridian with a given declination, and the sky will rotate overhead in a constant drift-scan. Each declination pointing will give us access to a $\sim6^{\circ}$ wide stripe of the sky. Because the cosmological signal is small, sample variance limits drive us to a total map sensitivity of $1-2\mu\mathrm{Jy}$. The fiducial design has a target noise temperature of 50\,K and a collecting area of $\sim29,000\,\mathrm{m}^{2}$. A few key design parameters for HIRAX-1024 are given in Table~\ref{tab:salient}. Assuming 50\,K system noise temperature, we will achieve a daily map depth of $\sim12\mu\mathrm{Jy}$\footnote{$\sigma_{\mathrm{K}} = T_{\mathrm{sys}} / \sqrt{N_{\mathrm{dish}} \times \delta\nu \times \Delta t}$, which can be converted to Jansky via [T/Jy]$ = 10^{-26} \times A_{\mathrm{e}} / 2k$} and will re-point every $\sim$150 days to achieve our final map sensitivity requirement. The complete survey of $\sim$15,000 square degrees can be accomplished in about four years.
\newline

We are currently building an 8-element prototype array (HIRAX-8) at the Hartebeesthoek Radio Astronomical Observatory (HartRAO) to develop the analog system and analysis pipelines. After initial decisions on instrumentation informed by the prototype, we will evaluate how to scale to a larger array with a 128-dish (HIRAX-128) instrument, and finally add dishes and a larger correlator for the full 1024-dish HIRAX array (HIRAX-1024). \newline

\subsection{Antennas: Dishes and Feeds}

The antennas for HIRAX are still under development and include the dishes and the feed with a choke structure. The primary instrument design consideration is to have fast mapping speed, requiring high sensitivity. This in turn drives the optical design to have a large collecting area and extremely efficient antennas: high aperture efficiency (60\%), low loss in the feed ($<0.15$\,dB), a low reflection coefficient at the feed ($<-15$\,dB), and low spillover to the ground ($<10$\,K). \newline

\begin{minipage}{0.35\textwidth}
  \centering
  \includegraphics[width=0.9\textwidth]{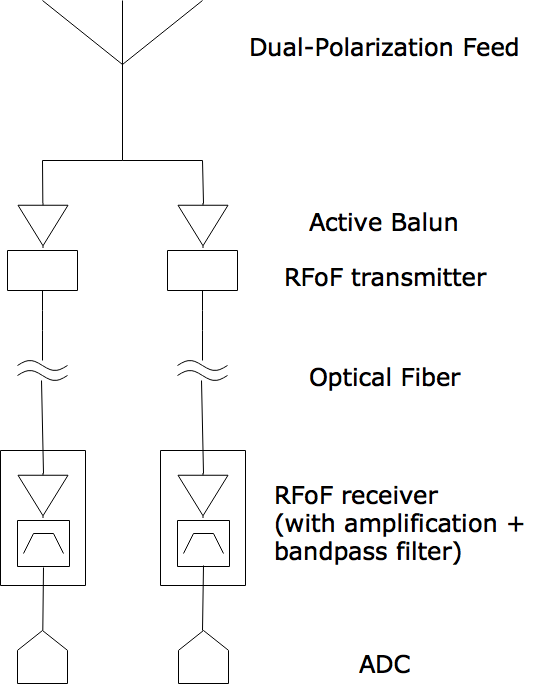} 
  \vspace{0.2in}
\end{minipage}
\begin{minipage}{0.5\textwidth}
  \centering
    \begin{tabular}{|l|l|} 
   \hline
   \rule[-1ex]{0pt}{3.5ex}  Frequency Range & 400--800\,MHz  \\ \hline
   \rule[-1ex]{0pt}{3.5ex}  Frequency Resolution & 390\,kHz, 1024 channels \\ \hline
   \rule[-1ex]{0pt}{3.5ex}  Dish size & 6\,m diameter, $f/D$=0.25 \\ \hline
   \rule[-1ex]{0pt}{3.5ex}  Interferometric layout & 32$\times$32 square grid, 7\,m spacing  \\ \hline
   \rule[-1ex]{0pt}{3.5ex}  Field of View & 15 deg$^{2}$--56 deg$^{2}$ \\ \hline
   \rule[-1ex]{0pt}{3.5ex}  Resolution & $\sim$5'--10'  \\ \hline 
   \rule[-1ex]{0pt}{3.5ex}  Beam Crossing Time & 17--32 minutes   \\ \hline  
   \rule[-1ex]{0pt}{3.5ex}  System Temperature & 50\,K \\ \hline
   \end{tabular}
   \vspace{0.2in}
  \captionof{table}{Table of instrumental parameters for HIRAX. }
       \label{tab:salient}
          \vspace{0.1in}
   \captionof{figure}{Shown is the signal chain for a single HIRAX dish. The signal is focused onto the dual-polarization antenna at the focus, and each linear polarization is amplified, transformed into an optical signal, carried on optical fiber, transformed back to RF, filtered, and amplified again before digitization.}
     \vspace{0.2in}
      \label{fig:instrument}
\end{minipage}

\begin{figure}[!tbp]
  \centering
  \begin{tabular}[c]{cc}
    \subfloat[]{\label{fig:dishpic}%
      \begin{tabular}[c]{c}
        \includegraphics[width=0.6\textwidth]{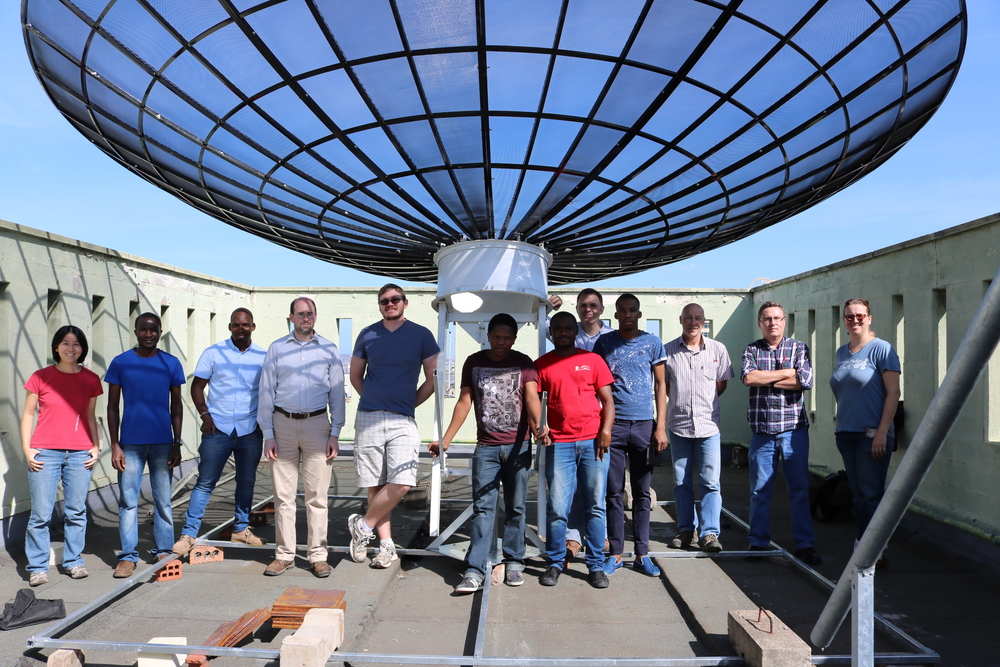}
      \end{tabular}}
    &
    \begin{tabular}[c]{c}
      \subfloat[]{\label{fig:activebalun}%
        \includegraphics[width=0.35\textwidth]{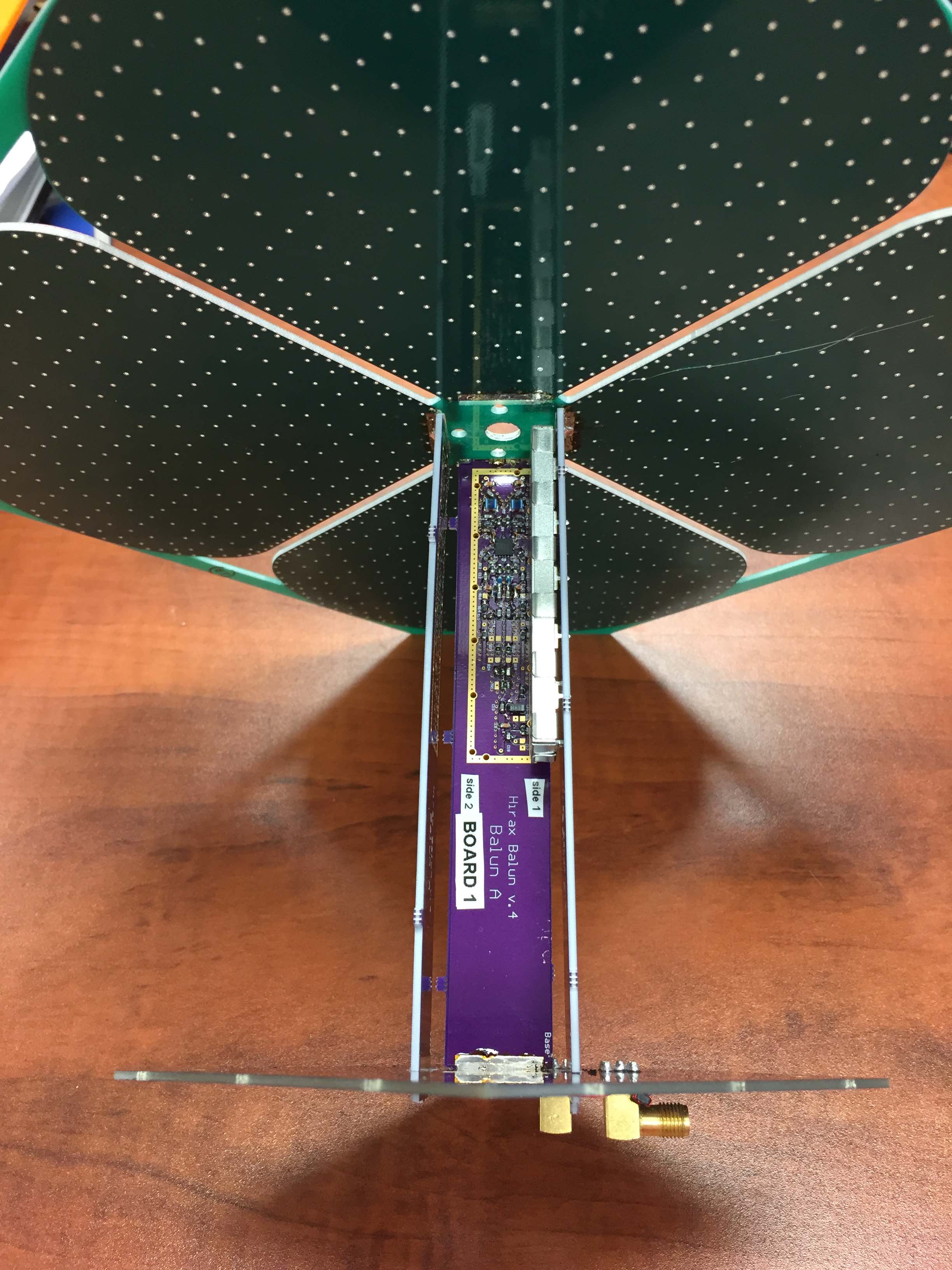}}
    \end{tabular}
  \end{tabular}
  \vspace{0.2in}
  \caption{(a) The first 6\,m prototype dish for HIRAX was assembled on a rooftop at Durban University of Technology in Durban, SA. (b) We are investigating the possibility of amplifying directly on the antenna balun to reduce system noise. This is a prototype with amplifier shown on the stem of the antenna. The amplification circuitry is protected from feed-back and oscillations with a small metal cover.}
  \label{fig:both}

\end{figure}

\textit{Dishes --} The dishes will be six meter diameter parabolic reflectors with $f/D = 0.25$. The small focal ratio will help reduce crosstalk between neighboring antennas.  Because we image the sky in strips by changing the center declination, the dishes must be able to tilt on one axis. We have initial dish designs for the HIRAX-8 prototype array, and two dishes have been built off-site to vet the design and assembly (see Figure~\ref{fig:dishpic}). We are in the process of optimizing the dishes and their mounting and rotation scheme based on experiences in the field and with the vendor. Any design must be cost efficient, easy to assemble, have a repeatable surface shape, and respect the tolerance for reflector surface imperfections ($\frac{\lambda}{50}$ = 7\,mm \cite{Ruze1952}). The dishes should also be rigid enough that the beam full-width-half-maximum does not change by more than 0.1\% \cite{2015PhRvD..91h3514S} upon tilting up to 25$^{\circ}$ to avoid calibrating the beams for every tilt. We would also like to minimize the reflections off of the support struts above the frame, for example by using a radio-transparent support. To further reduce ground spillover and cross-talk, we are considering adding reflective collars to the dishes. \newline

\textit{Feeds -- } The HIRAX feed will be based on the feed used for CHIME~\cite{6887670}, a dual-polarized clover-leaf shaped dipole antenna that was in turn based on a four-square antenna developed for Molonglo~\cite{Martin2008}. The CHIME feed has low-loss and small reflectivity characteristics across a wide band and is composed of: (i) a FR4-dielectric printed circuit board (PCB) which has four metalized curved petals to act as a wide-band dipole antenna, (ii) a low-loss teflon material \footnote{Rogers Arlon Diclad 880, dielectric constant of $\epsilon_{r} = 2.17$} balun for impedance matching, and (iii) a teflon support board. The signal current distribution, design parameters, and beam characteristics are described in detail in Refs.~\cite{2014SPIE.9145E..22B}~\cite{6887670}, here we note that the shape of the petals provides sensitivity to a wide bandwidth and there is one output signal for each linear polarization. \newline

The CHIME feed beam shape is broad and elliptical for each polarization, with an impedance chosen to minimize the noise of the low-noise amplifier at the feed output and is optimized for placement in a compact line-feed array. For HIRAX, we would like circular beams with good impedance matching to the dish and so are designing a new candidate feed which maintains the nice broad-band characteristics of the CHIME feed but is better matched to the HIRAX dishes. To circularize the beam and aid in reducing cross-talk and ground spillover, we will be adding a ring choke (in the form of a metallic can) that the feed will reside inside, which will also help us weather-proof the instrumentation at the focus. Various choke ring geometries are being simulated to optimize gain, reduce spillover, and reduce polarization artifacts. Optimization also includes choke size: while wider chokes are more effective at reducing spillover, they also increase the blockage of the center of the dish and reduce overall sensitivity. \newline

\subsection{Amplification}

To achieve fast integrated mapping speeds, we are targeting a system noise of 50\,K. In addition to minimizing losses in the optical chain as described above, this requires amplifying the signal either on or directly behind the feed with low noise amplifiers (LNAs). The gain specification is set by the required input level to the ADC: the averaged sky signal from all synchrotron emission is $\sim$35\,K and we need to digitize that signal such that its level on the input to the digitizer is -21\,dBm across the 400\,MHz bandwidth. The total input power from the average 35\,K sky and 50\,K system temperature would be -93\,dBm across the entire band, leading us to require $\sim$70\,dB of total gain. As noted below, 50\,dB of that gain must come before the Radio-Frequency over Fiber (RFoF) system for the system noise to be dominated by the LNA noise figure. We are investigating two alternatives for the amplification: including the amplification circuitry directly on the balun and backboard (a prototype of the feed with the active balun is shown in Figure~\ref{fig:activebalun}), and placing amplifiers at the SMA-connectorized outputs of the feeds. The primary benefit of amplifying directly on the feed is a reduction in system noise. Even with a low-loss material for the balun, the loss is $\sim$0.03\,dB\//inch, leading to an extra contribution of $\sim$12\,K to the system noise temperature, and so placing the LNA within the balun reduces the total noise temperature substantially. We are currently planning to use an Avago MGA-16116 GaAs MMIC LNA because it meets our noise specification. A 400-800\,MHz prototype LNA built with the device produced a gain of $\sim$18\,dB and a noise figure of $\sim$0.4\,dB (28\,K) while significantly reducing the expense of circuit board parts. \newline

To carry signals from the dishes to the correlator across the $\sim$250\,m$\times$250\,m HIRAX footprint we will use optical fiber. Optical fiber is an attractive solution for long cable runs when the loss (and its steep frequency dependence) of coaxial cable can be prohibitive. The RFoF modules were developed for radio telescopes and are composed of a transmitter at the dish that converts from RF to optical signals, an optical fiber, and a receiver at the correlator building to convert back to RF. They have relatively high noise (typically 27\,dB ENR), which sets the requirement of 50\,dB of gain before the RFoF transmitter-receiver pair for the system noise temperature to be dominated by a front-stage LNA. The RFoF receiver also contains the band-defining 400-800\,MHz filter and can be designed to have the final amplification stages required in the HIRAX signal chain. This system has been developed, tested in the lab, and a prototype set was deployed and tested on CHIME~\cite{2013JInst...810003M}. \newline      

\subsection{Digital Back End}

In an interferometer, the sky channels are correlated to form interferometric visibilities and these are the data products stored to disk. We require a correlator capable of processing 2048 spatial inputs across a 400\,MHz bandwidth with a flexible output to simultaneously record transient data. The HIRAX digital backend is an FX correlator and will leverage development for the CHIME correlator. The correlator architecture and implementation has been described thoroughly in Refs. ~\cite{2015arXiv150306202D}~\cite{2015arXiv150306203K}~\cite{2015arXiv150306189R} here we will describe the general data processing steps. The F-engine and corner-turn systems make use of the ICE custom FPGA electronics\footnote{http:\//\//www.mcgillcosmology.ca\//ice-system} developed for astronomy applications such as CHIME and the South Pole Telescope. The F-engine digitizes the input signals and performs frequency channelization using a set of custom boards, each of which takes 16 sky channels. For HIRAX-8 we can use just one board, for HIRAX-128 we will require 16 boards, and for HIRAX-1024 we will require 128 boards. The signal from each input is digitized at 8-bit precision at 800\,MHz. The signal is then sent through a customized poly-phase filter bank (PFB) and FFT algorithm, performed at 18+18-bit precision to channelize the signal into 1024 frequency channels between 400--800\,MHz~\cite{ICEboard}. The X-engine of the correlator will correlate all sky inputs for each of the 1024 frequency channels, and is implemented in an array of HPC GPU nodes. To network the data between the FPGA F-engine and the GPU X-engine, the data is sent via a custom backplane and point-to-point digital connections into the GPU nodes, during which it undergoes a corner-turn operation. The corner-turn takes the channelized samples from individual inputs, which are natively in 1024 frequency bins, and re-arranges and shuffles them into the format required for the X-engine, namely all sky inputs for a single frequency. \newline

The corner-turned data is sent to the GPU X-engine on 10\,Gbps lines. The diskless nodes will contain two network cards and two to four GPUs (depending on models selected), and the GPUs run a custom kernel to efficiently calculate the per-frequency correlation matrix. Each GPU correlator node is responsible for 32(4) frequency channels for all sky inputs for HIRAX-128 (HIRAX-1024). The data is accumulated and written to a separate storage system using standard UDP packets and the 10 Gigabit Ethernet protocol. \newline

The correlator will also form tied-array beams across the primary beam (`beamforming') to search for hydrogen absorbers, pulsars, and other transients. We chose to space the antennas on a regular grid to allow for efficient beam-forming algorithms, which can reduce the number of correlation operations from $\mathrm{N}^{2}$ (N is the number of spatial channels) to either NlogN for an efficient FFT algorithm or $\mathrm{N}^{3/2}$ for direct summation of similar baseline pairs.  The frequency resolution of the formed beams will be increased by a factor of 32, which can be done with a second-stage channelization within each polyphase filter bank frequency bin in the correlator. These formed beams will be searched for FRBs using tree algorithms~\cite{2015Natur.528..523M}, and a subset of approximately 20 beams will be passed to a pulsar search engine in HIRAX-1024.

\section{Instrument Characterization Plans}
\label{sec:chall}

The biggest challenge for all 21\,cm intensity mapping experiments is the presence of synchrotron foregrounds from our own Milky Way galaxy, which can be as bright as 700\,K in the HIRAX band. In theory these can be filtered due to their smooth spectral structure, as demonstrated in Ref.~\cite{2015PhRvD..91h3514S}. As described in that paper, this removal hinges on the precise calibration of the instrument, primarily its beam (to 0.1\%) and its gain (to 1\%). This foreground filtering degrades our ability to measure long modes along the line of sight because we filter smoothly in frequency. However, it does not impact our ability to measure the matter power spectrum, for which we are primarily interested in features on smaller scales. \newline

The current plan for calibrating the beam is to use a drone-based beam measurement technique. We will fly a broad-band antenna and source on a quadcopter drone in a pre-determined pattern to map out the near field beam at several elevations, allowing this information to be extrapolated to the far-field (or, when possible, we will measure the $\sim$100--200\,m far field directly, which is within range of commercially available drones). The current prototype drone calibration system uses a MAVRIK quadcopter\footnote{http:\//\//www.steadidrone.com}, produced by Steadidrone in South Africa. This drone is rated for up to 25 minutes of flight at minimum load, or a maximum payload of 2\,kg. It is capable of autonomous flight along preplanned flight paths using the ArduPilot autopilot software loaded on the onboard Pixhawk flight control module. The calibration source is a Valon 5009 dual frequency synthesizer module, which has a frequency range of 23\,MHz -- 6\,GHz, a 20\,MHz internal frequency reference, and a maximum output power of +15\,dBm. Using drones for beam mapping has been previously demonstrated \cite{2015PASP..127.1131C} and multiple groups are pursuing this method of beam calibration at a variety of wavelengths. We are currently building up the drone measurement program, with testing beginning this year.\newline

We have chosen to arrange the HIRAX dishes on a square grid to allow for maximum redundancy in baseline distances. Ultimately, this is to take advantage of redundancy for time-dependent gain calibration, as well as possible solutions for beam differences. The algorithm we are developing for HIRAX \cite{sievers} differs from traditional redundant calibration \cite{2010MNRAS.408.1029L} by using knowledge of the sky (whose prior can be incomplete and consist of only known bright point sources) to solve for gains only, using the expected visibility-visibility correlation function. This scheme allows the user to apply knowledge of the array non-idealities and partial sky knowledge to calibrate instrument gains and potentially extract beam shapes. \newline

\section{Current Status and Summary}

We are currently building HIRAX-8 at HartRAO, located $\sim$ 100\,km outside of Johannesburg. From an Radio-frequency interference (RFI) perspective, its proximity to Johannesburg is not ideal; however, it is easy to access and has many resources (e.g. facilities, staff, and infrastructure supporting the HartRAO 26\,m) which make it a useful site for building and debugging a small prototype array. HIRAX-8 will have 16 inputs (two polarizations from each of the 8 dishes), for which we can use one FPGA board and a single GPU node for correlation and data storage. We have taken test data with initial instrumentation on four small commercial dishes and a correlator FPGA board. We have ordered 8 prototype dishes to build HIRAX-8 and will use them to test instrument hardware until we have a stable design. When the design is finalized, we will build HIRAX-128 with 128 dishes (256 inputs) at an RFI-quiet location in South Africa. HIRAX-8 and HIRAX-128 are currently funded, and the latter will provide $\sim$2 times the collecting area as the CHIME pathfinder, nicely complementing it in the Southern hemisphere. 

HIRAX is a radio interferometer designed to measure 15,000 $\mathrm{deg}^{2}$ of the sky to a depth of $\sim1-2\mu\mathrm{Jy}$. The resulting maps of neutral hydrogen in 1024 frequency bins between 400--800\,MHz will provide low-resolution measurements of large-scale structure to better understand the nature of Dark Energy through its impact on the expansion rate of the universe. HIRAX will also monitor the sky for radio transients, both fast irregular bursts (such as FRBs) and pulsars. The prototpye array, HIRAX-8, will be constructed this year at HartRAO and be used to finalize the design for the full instrument.

\label{sec:conc}


\acknowledgments 
 
The DST-NRF-University of Kwa-Zulu Natal Flagship on \textit{HIRAX:Mapping the Southern Sky} is funded by the DST in partnership with UKZN and administered by the NRF. This work is based on research supported in part by the Department of Science and Technology and the National Research Foundation of South Africa as well as their South African Research Chairs Initiative. HCC, KM, JS, MH, and AW acknowledge support for their work which is based on research supported in part by the National Research Foundation of South Africa (Grant numbers 98772, 98957, 93565 and 91552). AW gratefully acknowledges support of the Pacyzynski Visiting Fellowship at Princeton University. BRS is funded by an SKA SA postdoctoral fellowship. The financial assistance of the South African SKA Project (SKA SA) towards this research is hereby acknowledged. Opinions expressed and conclusions arrived at are those of the author and are not necessarily to be attributed to the SKA SA (www.ska.ac.za).

\bibliography{bibliography} 
\bibliographystyle{spiebib} 

\end{document}